\documentclass[10pt,letterpaper]{article}
\usepackage{spconf,amsmath,amssymb,graphicx,booktabs,cite}
\usepackage[T1]{fontenc}
\usepackage{xurl,tikz,flushend}
\usetikzlibrary{arrows.meta,calc}
\usepackage[hidelinks]{hyperref}
\newcommand{\degree}{^{\circ}}
\newcommand{\RT}{\ensuremath{\mathrm{RT}_{60}}}
\title{\fontsize{14}{16}\selectfont{GAMF: Learned and Analytical Array Transfer Function Matching for Array-Generic Direction-of-Arrival Estimation}
}
\name{Zhiheng Jin$^{1,2,*}$,
Shichao Hu$^{1,2,*}$,
Chunyang Xu$^{1,2}$,
Mengyao Zhu$^{1,2,\dagger}$%
\thanks{$^*$These authors contributed equally and are listed in random order.}%
\thanks{\textsuperscript{\textdagger}Corresponding author: \texttt{zhu.mengyao@\allowbreak suda.edu.cn}.}%
}
\address{ $^1$Center of Auditory and Language Intelligence, Soochow University, Suzhou, China\\
$^2$School of Future Science and Engineering, Soochow University, Suzhou, China }
\begin{document}
\ninept
\maketitle
\raggedbottom
\begin{abstract}
Microphone positional encoding supports cross-array direction-of-arrival (DOA) estimation, but coordinates alone cannot fully describe device shadowing or microphone directivity. We propose a Generalizable ATF Matching Framework (GAMF) for DOA estimation across array geometries and microphone counts, using array transfer functions (ATFs) as acoustic descriptors. The learned branch incorporates ATF embeddings into geometry-conditioned neural estimation to match acoustic observations with candidate directions. The analytical branch performs normalized ATF matching adapted from generalized steered response power. A hybrid configuration combines their scores through adaptive gating. Simulations across array configurations show that both learned and hybrid configurations outperform a representative positional-encoding-based neural baseline, remain competitive with analytical ATF matching in clean, low-reverberation scenes, and substantially improve upon it under stronger noise or reverberation. On eight-microphone LOCATA Task~1 recordings, the hybrid configuration outperforms the evaluated state-of-the-art baselines for three-dimensional DOA estimation, achieving mean errors of $3.76^\circ$ for three-dimensional DOA and $2.87^\circ$ for azimuth.
\end{abstract}
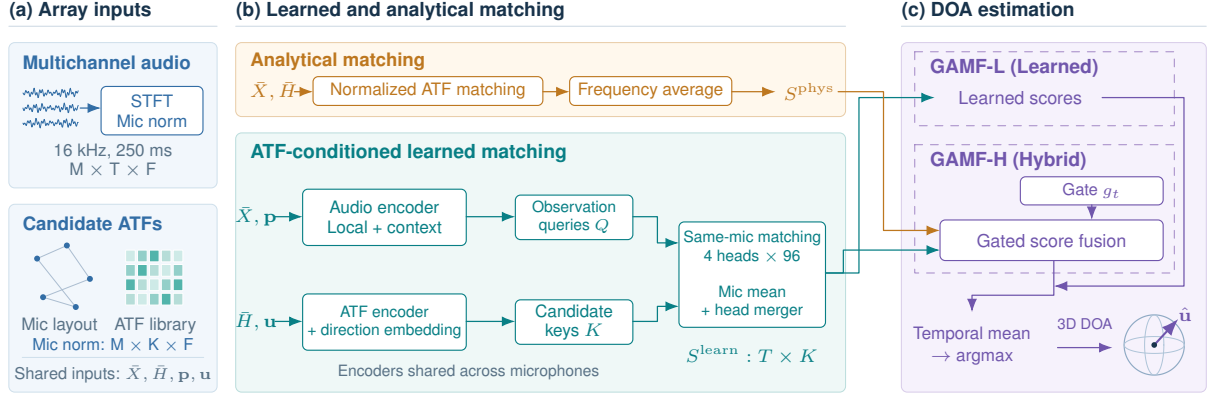
\begin{figure*}[t]
\centering
\resizebox{\textwidth}{!}{%
\definecolor{gamfColor18334B}{RGB}{24,51,75}
\definecolor{gamfColorCBD8E2}{RGB}{203,216,226}
\definecolor{gamfColorF2F7FC}{RGB}{242,247,252}
\definecolor{gamfColorCEDFEB}{RGB}{206,223,235}
\definecolor{gamfColorFFF8ED}{RGB}{255,248,237}
\definecolor{gamfColorE8CCA7}{RGB}{232,204,167}
\definecolor{gamfColorEFF8F7}{RGB}{239,248,247}
\definecolor{gamfColorBADDDC}{RGB}{186,221,220}
\definecolor{gamfColorF6F2FC}{RGB}{246,242,252}
\definecolor{gamfColorD8CCE9}{RGB}{216,204,233}
\definecolor{gamfColor326DA5}{RGB}{50,109,165}
\definecolor{gamfColorFFFFFF}{RGB}{255,255,255}
\definecolor{gamfColor597286}{RGB}{89,114,134}
\definecolor{gamfColorABBDD0}{RGB}{171,189,208}
\definecolor{gamfColorCEEAE7}{RGB}{206,234,231}
\definecolor{gamfColor93CFC9}{RGB}{147,207,201}
\definecolor{gamfColor52B4AB}{RGB}{82,180,171}
\definecolor{gamfColorB6DDD8}{RGB}{182,221,216}
\definecolor{gamfColorBD771F}{RGB}{189,119,31}
\definecolor{gamfColor087F85}{RGB}{8,127,133}
\definecolor{gamfColorA68CC8}{RGB}{166,140,200}
\definecolor{gamfColor7056AC}{RGB}{112,86,172}
\definecolor{gamfColor8FAAC4}{RGB}{143,170,196}
\definecolor{gamfColorAEC4D8}{RGB}{174,196,216}
\begin{tikzpicture}[x=0.01cm,y=-0.01cm,>=Latex,line cap=round,line join=round]
\node[anchor=west,align=left,text=gamfColor18334B,font=\sffamily\bfseries\fontsize{7.38}{9.01}\selectfont,inner sep=0pt] at (20,26.4) {(a) Array inputs};
\draw[draw=gamfColorCBD8E2,line width=0.31pt] (20,46.64) -- (300,46.64);
\node[anchor=west,align=left,text=gamfColor18334B,font=\sffamily\bfseries\fontsize{7.38}{9.01}\selectfont,inner sep=0pt] at (325,26.4) {(b) Learned and analytical matching};
\draw[draw=gamfColorCBD8E2,line width=0.31pt] (325,46.64) -- (1145,46.64);
\node[anchor=west,align=left,text=gamfColor18334B,font=\sffamily\bfseries\fontsize{7.38}{9.01}\selectfont,inner sep=0pt] at (1220,26.4) {(c) DOA estimation};
\draw[draw=gamfColorCBD8E2,line width=0.31pt] (1220,46.64) -- (1630,46.64);
\path[fill=gamfColorF2F7FC,draw=gamfColorCEDFEB,line width=0.37pt,rounded corners=2.27pt] (20,68.64) rectangle (300,264.0);
\path[fill=gamfColorF2F7FC,draw=gamfColorCEDFEB,line width=0.37pt,rounded corners=2.27pt] (20,286.0) rectangle (300,538.56);
\path[fill=gamfColorFFF8ED,draw=gamfColorE8CCA7,line width=0.37pt,rounded corners=2.27pt] (325,68.64) rectangle (1145,168.07999999999998);
\path[fill=gamfColorEFF8F7,draw=gamfColorBADDDC,line width=0.37pt,rounded corners=2.27pt] (325,190.08) rectangle (1145,538.5600000000001);
\path[fill=gamfColorF6F2FC,draw=gamfColorD8CCE9,line width=0.37pt,rounded corners=2.27pt] (1220,68.64) rectangle (1630,538.5600000000001);
\node[anchor=west,align=left,text=gamfColor326DA5,font=\sffamily\bfseries\fontsize{7.10}{8.66}\selectfont,inner sep=0pt] at (38,95.04) {Multichannel audio};
\draw[draw=gamfColor326DA5,line width=0.28pt] (39.0,136.4) -- (39.75,137.4265512325528) -- (40.5,138.87516116525904) -- (41.25,138.09556421191107) -- (42.0,135.1654559097795) -- (42.75,133.07541208047246) -- (43.5,133.9156210489924) -- (44.25,136.0846839824849) -- (45.0,136.45775014815194) -- (45.75,134.79771551115178) -- (46.5,134.283283315109) -- (47.25,137.03927338931595) -- (48.0,140.6886037937731) -- (48.75,141.08655804833523) -- (49.5,137.91205163379047) -- (50.25,135.29135457081287) -- (51.0,136.15201976643067) -- (51.75,138.04052340959475) -- (52.5,136.60160690268813) -- (53.25,132.1490049414419) -- (54.0,129.87998354952148) -- (54.75,132.93035109627758) -- (55.5,137.86125681806098) -- (56.25,139.1846889855988) -- (57.0,136.80494480017217) -- (57.75,135.86414291059558) -- (58.5,139.1154756615454) -- (59.25,142.66682725893165) -- (60.0,141.27149492787282) -- (60.75,135.7591101726313) -- (61.5,132.09733743456286) -- (62.25,133.4123369146846) -- (63.0,136.15940547565148) -- (63.75,135.5820303298735) -- (64.5,132.50042679503068) -- (65.25,131.93487984407133) -- (66.0,135.83251246359058) -- (66.75,140.25078793337642) -- (67.5,140.69187870486476) -- (68.25,137.89670198067432) -- (69.0,136.2673754599082) -- (69.75,137.62729173680955) -- (70.5,139.08944337162805) -- (71.25,137.5250115086483) -- (72.0,134.03851202815684) -- (72.75,132.4180581035631) -- (73.5,134.08560095014053) -- (74.25,136.49472049813087) -- (75.0,136.7858016072084) -- (75.75,135.47964066476928) -- (76.5,135.39168878535804) -- (77.25,137.55559563879484) -- (78.0,139.7174918969627) -- (78.75,139.2997440958078) -- (79.5,136.80886990801974) -- (80.25,135.16612210363354) -- (81.0,135.84527491283976) -- (81.75,136.96004827899336) -- (82.5,135.95567264874208) -- (83.25,133.3841720199338) -- (84.0,132.51754634746277) -- (84.75,134.9875571504834) -- (85.5,138.29313067466262) -- (86.25,138.90906037717323) -- (87.0,137.0529017375809) -- (87.75,136.4014889777079) -- (88.5,138.75610269474572) -- (89.25,141.04024946065334) -- (90.0,139.28921245301933) -- (90.75,134.4406278521417) -- (91.5,131.60597628067805) -- (92.25,133.33262557852092) -- (93.0,136.15972317353237) -- (93.75,135.58704656038444) -- (94.5,132.65039669321294) -- (95.25,132.62667660296148) -- (96.0,137.3949025892721) -- (96.75,142.2590658000471) -- (97.5,142.01336309061136) -- (98.25,137.9053536749263) -- (99.0,135.69954316061805) -- (99.75,137.5781259289775) -- (100.5,139.32335022454444) -- (101.25,136.6583492403764) -- (102.0,131.6292012173375) -- (102.75,129.99822995746732) -- (103.5,133.4330316328524) -- (104.25,137.44853273677796) -- (105.0,137.70017776701192) -- (105.75,135.60742291649302) -- (106.5,135.71060093191667) -- (107.25,138.96171062344604) -- (108.0,141.4297190998153) -- (108.75,139.84522039617497) -- (109.5,135.99392851613464) -- (110.25,134.11478524618153) -- (111.0,135.31311728744652) -- (111.75,136.63125034761384) -- (112.5,135.54450385410098) -- (113.25,133.41125108824602);
\draw[draw=gamfColor326DA5,line width=0.28pt] (39.0,156.64000000000001) -- (39.75,159.2961021949619) -- (40.5,160.0827201863979) -- (41.25,158.16068484613066) -- (42.0,155.96175696468026) -- (42.75,155.95850991728065) -- (43.5,157.3215802258293) -- (44.25,157.0774540926201) -- (45.0,154.42101493457028) -- (45.75,152.28162916149267) -- (46.5,153.61633591085075) -- (47.25,157.2271686978345) -- (48.0,159.07406667629934) -- (48.75,157.6985211817017) -- (49.5,156.27731549701878) -- (50.25,158.0581408597868) -- (51.0,161.3349366727941) -- (51.75,161.3458728642972) -- (52.5,156.93591513869947) -- (53.25,152.64172163379135) -- (54.0,152.8207669767292) -- (54.75,155.78505292964485) -- (55.5,156.33791071363956) -- (56.25,153.35230600892316) -- (57.0,151.60588642393304) -- (57.75,155.0132425325067) -- (58.5,160.73769031405882) -- (59.25,162.67537020405805) -- (60.0,159.582466844401) -- (60.75,156.46911063059596) -- (61.5,157.39231609870913) -- (62.25,159.8978525363887) -- (63.0,158.8481814965188) -- (63.75,154.00124256461837) -- (64.5,150.58650812517047) -- (65.25,152.23705488622053) -- (66.0,156.28683902087985) -- (66.75,157.7436287898136) -- (67.5,156.02147712558101) -- (68.25,155.13209619344485) -- (69.0,157.59800545264338) -- (69.75,160.86718799573177) -- (70.5,160.93770854590355) -- (71.25,157.77583079394387) -- (72.0,155.12697685887406) -- (72.75,155.4203540297151) -- (73.5,156.90486338489296) -- (74.25,156.5518273399818) -- (75.0,154.35668778073477) -- (75.75,153.2366710839185) -- (76.5,154.98013462874712) -- (77.25,157.83457701887056) -- (78.0,158.82684520797855) -- (78.75,157.58460669437997) -- (79.5,156.604939765305) -- (80.25,157.70093078263488) -- (81.0,159.4049800571714) -- (81.75,158.8779751959393) -- (82.5,155.88589170608796) -- (83.25,153.4478455760272) -- (84.0,154.00855767043527) -- (84.75,156.1527030114638) -- (85.5,156.53026629985598) -- (86.25,154.648495507391) -- (87.0,153.84399445400683) -- (87.75,156.64648047542389) -- (88.5,160.7662888776956) -- (89.25,161.66941802637882) -- (90.0,158.65234137839192) -- (90.75,155.9266733765245) -- (91.5,156.78870823062906) -- (92.25,158.88529624803755) -- (93.0,157.53745276723262) -- (93.75,152.74934753701032) -- (94.5,149.85726367016036) -- (95.25,152.45165714441845) -- (96.0,157.3517293825601) -- (96.75,158.88368800781348) -- (97.5,156.61275494466508) -- (98.25,155.62514390661957) -- (99.0,158.9486831332852) -- (99.75,162.9139643196928) -- (100.5,162.10783291998837) -- (101.25,156.9337127674231) -- (102.0,153.1362272177372) -- (102.75,154.0685675371694) -- (103.5,156.57890950549097) -- (104.25,155.99008570398743) -- (105.0,152.82323270921003) -- (105.75,151.8718535643518) -- (106.5,155.30516919013343) -- (107.25,159.62340377130207) -- (108.0,160.4306775286144) -- (108.75,158.08793890755473) -- (109.5,156.62157888939913) -- (110.25,157.93946820658414) -- (111.0,159.50307979547426) -- (111.75,158.243138147813) -- (112.5,154.9230709996136) -- (113.25,153.07368055110678);
\draw[draw=gamfColor326DA5,line width=0.28pt] (39.0,176.88) -- (39.75,178.72364504856588) -- (40.5,178.12505814508012) -- (41.25,176.82769484581522) -- (42.0,177.38163153837598) -- (42.75,179.4681665932885) -- (43.5,180.10089768630698) -- (44.25,177.66803092742327) -- (45.0,174.4244043567708) -- (45.75,173.7726088611003) -- (46.5,175.72933132581497) -- (47.25,176.8752238134312) -- (48.0,175.22165988190957) -- (48.75,173.33728482223216) -- (49.5,174.9760298196827) -- (50.25,179.5210949823644) -- (51.0,182.20135045400042) -- (51.75,180.32466450980942) -- (52.5,176.9981603608731) -- (53.25,176.81043701702765) -- (54.0,179.2729356322366) -- (54.75,179.42578915670623) -- (55.5,175.09230410594182) -- (56.25,170.54261372566594) -- (57.0,171.03516885352568) -- (57.75,175.65797546785473) -- (58.5,178.59250738929367) -- (59.25,177.13502161710923) -- (60.0,175.188148314068) -- (60.75,177.336225986694) -- (61.5,181.99561881118566) -- (62.25,183.38811356049376) -- (63.0,179.5067656330375) -- (63.75,174.84651621619972) -- (64.5,174.23814196792046) -- (65.25,176.58727736075684) -- (66.0,177.06586015368688) -- (66.75,174.22179842654108) -- (67.5,171.91974262457373) -- (68.25,173.7538502119075) -- (69.0,178.04784965028674) -- (69.75,180.22062710605442) -- (70.5,178.83468030297368) -- (71.25,176.98237248543953) -- (72.0,177.60650818787872) -- (72.75,179.54398683626144) -- (73.5,179.48061164505486) -- (74.25,176.68999971882442) -- (75.0,174.0268406786198) -- (75.75,174.12270641325765) -- (76.5,176.09465303960476) -- (77.25,177.01526979687085) -- (78.0,175.92562311993228) -- (78.75,175.00100225441616) -- (79.5,176.4332438406803) -- (80.25,179.26032459281467) -- (81.0,180.42257528829848) -- (81.75,178.7383180386901) -- (82.5,176.50943445110795) -- (83.25,176.44637118814683) -- (84.0,177.91890493569153) -- (84.75,177.7658674764192) -- (85.5,174.86829058289885) -- (86.25,172.21891068381996) -- (87.0,173.20572177497866) -- (87.75,176.88551385392137) -- (88.5,178.98278409584822) -- (89.25,177.6745628530012) -- (90.0,176.16533292085873) -- (90.75,178.06854810885645) -- (91.5,181.83471851914285) -- (92.25,182.37365190182265) -- (93.0,178.09006842555473) -- (93.75,173.4886964454458) -- (94.5,173.30014714857114) -- (95.25,176.12738079176435) -- (96.0,176.6541954638306) -- (96.75,173.44547380849355) -- (97.5,171.23719577694672) -- (98.25,174.27798815031815) -- (99.0,180.07523048024768) -- (99.75,182.48154884875143) -- (100.5,179.86521524499844) -- (101.25,176.9390381306268) -- (102.0,177.86460575567327) -- (102.75,180.50306826442957) -- (103.5,179.78095369704792) -- (104.25,175.12916687771752) -- (105.0,171.45540589663625) -- (105.75,172.52009605571166) -- (106.5,176.12697473905385) -- (107.25,177.5421692504944) -- (108.0,175.94650451921055) -- (108.75,174.99942906484108) -- (109.5,177.26616554679674) -- (110.25,180.569426090658) -- (111.0,181.06073994331192) -- (111.75,178.38110812816325) -- (112.5,175.88017475005773) -- (113.25,176.0149676683552);
\path[fill=gamfColorFFFFFF,draw=gamfColor326DA5,line width=0.43pt,rounded corners=1.99pt] (144,127.6) rectangle (281,192.72);
\node[anchor=center,align=center,text=gamfColor326DA5,font=\sffamily\fontsize{6.53}{7.97}\selectfont,inner sep=0pt] at (212.5,160.16) {STFT\\Mic norm};
\draw[->,draw=gamfColor326DA5,line width=0.48pt] (116,156.64000000000001) -- (144,156.64000000000001);
\node[anchor=center,align=center,text=gamfColor597286,font=\sffamily\fontsize{6.53}{7.97}\selectfont,inner sep=0pt] at (160,227.04) {16 kHz, 250 ms\\M $\times$ T $\times$ F};
\node[anchor=west,align=left,text=gamfColor326DA5,font=\sffamily\bfseries\fontsize{7.10}{8.66}\selectfont,inner sep=0pt] at (38,311.52) {Candidate ATFs};
\draw[draw=gamfColorABBDD0,line width=0.28pt] (58,359.04) -- (98,341.44);
\draw[draw=gamfColorABBDD0,line width=0.28pt] (98,341.44) -- (129,386.32);
\draw[draw=gamfColorABBDD0,line width=0.28pt] (129,386.32) -- (66,410.96);
\draw[draw=gamfColorABBDD0,line width=0.28pt] (66,410.96) -- (110,424.16);
\draw[draw=gamfColorABBDD0,line width=0.28pt] (110,424.16) -- (58,359.04);
\fill[fill=gamfColor326DA5] (58,359.04) circle[radius=0.04cm];
\fill[fill=gamfColor326DA5] (98,341.44) circle[radius=0.04cm];
\fill[fill=gamfColor326DA5] (129,386.32) circle[radius=0.04cm];
\fill[fill=gamfColor326DA5] (66,410.96) circle[radius=0.04cm];
\fill[fill=gamfColor326DA5] (110,424.16) circle[radius=0.04cm];
\path[fill=gamfColorCEEAE7,draw=gamfColorFFFFFF,line width=0.14pt,rounded corners=0.28pt] (177,347.6) rectangle (190,363.44);
\path[fill=gamfColor93CFC9,draw=gamfColorFFFFFF,line width=0.14pt,rounded corners=0.28pt] (193,347.6) rectangle (206,363.44);
\path[fill=gamfColor52B4AB,draw=gamfColorFFFFFF,line width=0.14pt,rounded corners=0.28pt] (209,347.6) rectangle (222,363.44);
\path[fill=gamfColorB6DDD8,draw=gamfColorFFFFFF,line width=0.14pt,rounded corners=0.28pt] (225,347.6) rectangle (238,363.44);
\path[fill=gamfColorCEEAE7,draw=gamfColorFFFFFF,line width=0.14pt,rounded corners=0.28pt] (241,347.6) rectangle (254,363.44);
\path[fill=gamfColor93CFC9,draw=gamfColorFFFFFF,line width=0.14pt,rounded corners=0.28pt] (177,366.96) rectangle (190,382.79999999999995);
\path[fill=gamfColor52B4AB,draw=gamfColorFFFFFF,line width=0.14pt,rounded corners=0.28pt] (193,366.96) rectangle (206,382.79999999999995);
\path[fill=gamfColorB6DDD8,draw=gamfColorFFFFFF,line width=0.14pt,rounded corners=0.28pt] (209,366.96) rectangle (222,382.79999999999995);
\path[fill=gamfColorCEEAE7,draw=gamfColorFFFFFF,line width=0.14pt,rounded corners=0.28pt] (225,366.96) rectangle (238,382.79999999999995);
\path[fill=gamfColor93CFC9,draw=gamfColorFFFFFF,line width=0.14pt,rounded corners=0.28pt] (241,366.96) rectangle (254,382.79999999999995);
\path[fill=gamfColor52B4AB,draw=gamfColorFFFFFF,line width=0.14pt,rounded corners=0.28pt] (177,386.32) rectangle (190,402.15999999999997);
\path[fill=gamfColorB6DDD8,draw=gamfColorFFFFFF,line width=0.14pt,rounded corners=0.28pt] (193,386.32) rectangle (206,402.15999999999997);
\path[fill=gamfColorCEEAE7,draw=gamfColorFFFFFF,line width=0.14pt,rounded corners=0.28pt] (209,386.32) rectangle (222,402.15999999999997);
\path[fill=gamfColor93CFC9,draw=gamfColorFFFFFF,line width=0.14pt,rounded corners=0.28pt] (225,386.32) rectangle (238,402.15999999999997);
\path[fill=gamfColor52B4AB,draw=gamfColorFFFFFF,line width=0.14pt,rounded corners=0.28pt] (241,386.32) rectangle (254,402.15999999999997);
\path[fill=gamfColorB6DDD8,draw=gamfColorFFFFFF,line width=0.14pt,rounded corners=0.28pt] (177,405.68) rectangle (190,421.52);
\path[fill=gamfColorCEEAE7,draw=gamfColorFFFFFF,line width=0.14pt,rounded corners=0.28pt] (193,405.68) rectangle (206,421.52);
\path[fill=gamfColor93CFC9,draw=gamfColorFFFFFF,line width=0.14pt,rounded corners=0.28pt] (209,405.68) rectangle (222,421.52);
\path[fill=gamfColor52B4AB,draw=gamfColorFFFFFF,line width=0.14pt,rounded corners=0.28pt] (225,405.68) rectangle (238,421.52);
\path[fill=gamfColorB6DDD8,draw=gamfColorFFFFFF,line width=0.14pt,rounded corners=0.28pt] (241,405.68) rectangle (254,421.52);
\node[anchor=center,align=center,text=gamfColor597286,font=\sffamily\fontsize{6.53}{7.97}\selectfont,inner sep=0pt] at (88,448.8) {Mic layout};
\node[anchor=center,align=center,text=gamfColor597286,font=\sffamily\fontsize{6.53}{7.97}\selectfont,inner sep=0pt] at (217,448.8) {ATF library};
\node[anchor=center,align=center,text=gamfColor326DA5,font=\sffamily\fontsize{6.25}{7.62}\selectfont,inner sep=0pt] at (160,476.08) {Mic norm: M $\times$ K $\times$ F};
\draw[draw=gamfColorCEDFEB,line width=0.28pt] (38,491.92) -- (282,491.92);
\node[anchor=center,align=center,text=gamfColor597286,font=\sffamily\fontsize{6.25}{7.62}\selectfont,inner sep=0pt] at (160,513.92) {Shared inputs: $\bar X,\bar H,\mathbf p,\mathbf u$};
\node[anchor=west,align=left,text=gamfColorBD771F,font=\sffamily\bfseries\fontsize{7.10}{8.66}\selectfont,inner sep=0pt] at (344,93.28) {Analytical matching};
\node[anchor=center,align=center,text=gamfColorBD771F,font=\sffamily\fontsize{6.53}{7.97}\selectfont,inner sep=0pt] at (376,134.64000000000001) {$\bar X,\bar H$};
\path[fill=gamfColorFFFFFF,draw=gamfColorBD771F,line width=0.43pt,rounded corners=1.99pt] (429,114.4) rectangle (738,153.12);
\node[anchor=center,align=center,text=gamfColorBD771F,font=\sffamily\fontsize{6.53}{7.97}\selectfont,inner sep=0pt] at (583.5,133.76) {Normalized ATF matching};
\path[fill=gamfColorFFFFFF,draw=gamfColorBD771F,line width=0.43pt,rounded corners=1.99pt] (774,114.4) rectangle (992,153.12);
\node[anchor=center,align=center,text=gamfColorBD771F,font=\sffamily\fontsize{6.53}{7.97}\selectfont,inner sep=0pt] at (883.0,133.76) {Frequency average};
\draw[->,draw=gamfColorBD771F,line width=0.48pt] (404,134.64000000000001) -- (429,134.64000000000001);
\draw[->,draw=gamfColorBD771F,line width=0.48pt] (738,134.64000000000001) -- (774,134.64000000000001);
\draw[->,draw=gamfColorBD771F,line width=0.48pt] (992,134.64000000000001) -- (1051,134.64000000000001);
\node[anchor=center,align=center,text=gamfColorBD771F,font=\sffamily\fontsize{7.10}{8.66}\selectfont,inner sep=0pt] at (1096,134.64000000000001) {$S^{\rm phys}$};
\node[anchor=west,align=left,text=gamfColor087F85,font=\sffamily\bfseries\fontsize{7.38}{9.01}\selectfont,inner sep=0pt] at (344,217.36) {ATF-conditioned learned matching};
\node[anchor=center,align=center,text=gamfColor087F85,font=\sffamily\fontsize{6.82}{8.32}\selectfont,inner sep=0pt] at (352,302.72) {$\bar X,\mathbf p$};
\path[fill=gamfColorFFFFFF,draw=gamfColor087F85,line width=0.43pt,rounded corners=1.99pt] (415,265.76) rectangle (635,341.44);
\node[anchor=center,align=center,text=gamfColor087F85,font=\sffamily\fontsize{6.53}{7.97}\selectfont,inner sep=0pt] at (525.0,303.6) {Audio encoder\\Local + context};
\draw[->,draw=gamfColor087F85,line width=0.48pt] (378,302.72) -- (415,302.72);
\path[fill=gamfColorFFFFFF,draw=gamfColor087F85,line width=0.43pt,rounded corners=1.99pt] (701,273.68) rectangle (859,333.52);
\node[anchor=center,align=center,text=gamfColor087F85,font=\sffamily\fontsize{5.96}{7.28}\selectfont,inner sep=0pt] at (780.0,303.6) {Observation\\queries $Q$};
\draw[->,draw=gamfColor087F85,line width=0.48pt] (635,302.72) -- (701,302.72);
\node[anchor=center,align=center,text=gamfColor087F85,font=\sffamily\fontsize{6.82}{8.32}\selectfont,inner sep=0pt] at (352,443.52) {$\bar H,\mathbf u$};
\path[fill=gamfColorFFFFFF,draw=gamfColor087F85,line width=0.43pt,rounded corners=1.99pt] (415,406.56) rectangle (635,482.24);
\node[anchor=center,align=center,text=gamfColor087F85,font=\sffamily\fontsize{5.96}{7.28}\selectfont,inner sep=0pt] at (525.0,444.4) {ATF encoder\\+ direction embedding};
\draw[->,draw=gamfColor087F85,line width=0.48pt] (378,443.52) -- (415,443.52);
\path[fill=gamfColorFFFFFF,draw=gamfColor087F85,line width=0.43pt,rounded corners=1.99pt] (701,414.48) rectangle (859,474.32000000000005);
\node[anchor=center,align=center,text=gamfColor087F85,font=\sffamily\fontsize{6.53}{7.97}\selectfont,inner sep=0pt] at (780.0,444.4) {Candidate\\keys $K$};
\draw[->,draw=gamfColor087F85,line width=0.48pt] (635,443.52) -- (701,443.52);
\path[fill=gamfColorFFFFFF,draw=gamfColor087F85,line width=0.43pt,rounded corners=1.99pt] (921,309.76) rectangle (1118,451.44);
\node[anchor=center,align=center,text=gamfColor087F85,font=\sffamily\fontsize{5.96}{7.28}\selectfont,inner sep=0pt] at (1019.5,380.6) {Same-mic matching\\4 heads $\times$ 96\\\\Mic mean\\+ head merger};
\draw[->,draw=gamfColor087F85,line width=0.48pt] (859,302.72) -- (892,302.72) -- (892,337.92) -- (921,337.92);
\draw[->,draw=gamfColor087F85,line width=0.48pt] (859,443.52) -- (892,443.52) -- (892,423.28000000000003) -- (921,423.28000000000003);
\node[anchor=center,align=center,text=gamfColor087F85,font=\sffamily\fontsize{6.82}{8.32}\selectfont,inner sep=0pt] at (1019,488.4) {$S^{\rm learn}:T\times K$};
\node[anchor=center,align=center,text=gamfColor597286,font=\sffamily\fontsize{5.96}{7.28}\selectfont,inner sep=0pt] at (638,512.16) {Encoders shared across microphones};
\draw[dashed,draw=gamfColorA68CC8,line width=0.40pt] (1238,80.96) -- (1587,80.96) -- (1587,180.4) -- (1238,180.4) -- (1238,80.96);
\draw[dashed,draw=gamfColorA68CC8,line width=0.40pt] (1238,205.92) -- (1587,205.92) -- (1587,378.4) -- (1238,378.4) -- (1238,205.92);
\node[anchor=west,align=left,text=gamfColor7056AC,font=\sffamily\bfseries\fontsize{7.38}{9.01}\selectfont,inner sep=0pt] at (1258,102.08) {GAMF-L (Learned)};
\node[anchor=center,align=center,text=gamfColor7056AC,font=\sffamily\fontsize{6.82}{8.32}\selectfont,inner sep=0pt] at (1380,142.56) {Learned scores};
\draw[->,draw=gamfColor087F85,line width=0.48pt] (1118,378.4) -- (1163,378.4) -- (1163,142.56) -- (1265,142.56);
\draw[->,draw=gamfColor7056AC,line width=0.48pt] (1490,142.56) -- (1600,142.56) -- (1600,396.0) -- (1425,396.0);
\node[anchor=west,align=left,text=gamfColor7056AC,font=\sffamily\bfseries\fontsize{7.38}{9.01}\selectfont,inner sep=0pt] at (1258,227.92) {GAMF-H (Hybrid)};
\path[fill=gamfColorFFFFFF,draw=gamfColor7056AC,line width=0.43pt,rounded corners=1.99pt] (1380,249.04) rectangle (1572,286.88);
\node[anchor=center,align=center,text=gamfColor7056AC,font=\sffamily\fontsize{6.25}{7.62}\selectfont,inner sep=0pt] at (1476.0,267.96) {Gate $g_t$};
\draw[->,draw=gamfColor7056AC,line width=0.48pt] (1476,286.88) -- (1476,307.12);
\path[fill=gamfColorFFFFFF,draw=gamfColor7056AC,line width=0.43pt,rounded corners=1.99pt] (1272,307.12) rectangle (1572,361.68);
\node[anchor=center,align=center,text=gamfColor7056AC,font=\sffamily\fontsize{6.82}{8.32}\selectfont,inner sep=0pt] at (1422.0,334.4) {Gated score fusion};
\draw[->,draw=gamfColorBD771F,line width=0.48pt] (1135,134.64000000000001) -- (1196,134.64000000000001) -- (1196,321.2) -- (1272,321.2);
\draw[->,draw=gamfColor087F85,line width=0.48pt] (1118,378.4) -- (1142,378.4) -- (1142,347.6) -- (1272,347.6);
\draw[->,draw=gamfColor7056AC,line width=0.48pt] (1425,361.68) -- (1425,408.32) -- (1315,408.32) -- (1315,431.2);
\node[anchor=center,align=center,text=gamfColor7056AC,font=\sffamily\fontsize{6.53}{7.97}\selectfont,inner sep=0pt] at (1315,462.0) {Temporal mean};
\node[anchor=center,align=center,text=gamfColor7056AC,font=\sffamily\fontsize{6.53}{7.97}\selectfont,inner sep=0pt] at (1315,493.68) {$\to$ argmax};
\draw[->,draw=gamfColor7056AC,line width=0.48pt] (1430,478.72) -- (1504,478.72);
\node[anchor=center,align=center,text=gamfColor7056AC,font=\sffamily\fontsize{5.68}{6.93}\selectfont,inner sep=0pt] at (1467,446.16) {3D DOA};
\draw[draw=gamfColor8FAAC4,line width=0.45pt] (1600.48,475.2) -- (1600.4245235668652,477.31855950871443) -- (1600.2582463245078,479.43131219307463) -- (1599.9816240272912,481.53246714482856) -- (1599.5954148777043,483.61626524430295) -- (1599.1006774481814,485.6769949457501) -- (1598.4987677796278,487.70900793229794) -- (1597.7913356646068,489.7067345975938) -- (1596.9803201253724,491.6646993117084) -- (1596.067944099145,493.5775354294568) -- (1595.056708345194,495.44) -- (1593.9493845904308,497.2469881374083) -- (1592.749007932298,498.9935470127993) -- (1591.4588685197782,500.6748894296975) -- (1590.082502535325,502.28640694540655) -- (1588.6236825024314,503.8236825024315) -- (1587.0864069454067,505.28250253532497) -- (1585.4748894296974,506.65886851977814) -- (1583.7935470127993,507.94900793229795) -- (1582.0469881374083,509.1493845904307) -- (1580.24,510.2567083451941) -- (1578.3775354294569,511.26794409914515) -- (1576.4646993117085,512.1803201253724) -- (1574.5067345975938,512.9913356646067) -- (1572.509007932298,513.6987677796278) -- (1570.47699494575,514.3006774481814) -- (1568.4162652443028,514.7954148777045) -- (1566.3324671448286,515.1816240272912) -- (1564.2313121930747,515.4582463245076) -- (1562.1185595087143,515.6245235668652) -- (1560.0,515.68) -- (1557.8814404912857,515.6245235668652) -- (1555.7686878069253,515.4582463245076) -- (1553.6675328551714,515.1816240272912) -- (1551.5837347556972,514.7954148777045) -- (1549.52300505425,514.3006774481814) -- (1547.490992067702,513.6987677796279) -- (1545.4932654024062,512.9913356646067) -- (1543.5353006882915,512.1803201253724) -- (1541.6224645705431,511.26794409914515) -- (1539.76,510.2567083451941) -- (1537.9530118625917,509.1493845904307) -- (1536.2064529872007,507.94900793229795) -- (1534.5251105703026,506.65886851977814) -- (1532.9135930545935,505.28250253532497) -- (1531.3763174975686,503.8236825024315) -- (1529.917497464675,502.28640694540655) -- (1528.5411314802218,500.6748894296975) -- (1527.250992067702,498.9935470127993) -- (1526.0506154095692,497.2469881374083) -- (1524.943291654806,495.44) -- (1523.932055900855,493.57753542945693) -- (1523.0196798746274,491.6646993117084) -- (1522.2086643353932,489.7067345975938) -- (1521.5012322203722,487.70900793229794) -- (1520.8993225518186,485.6769949457501) -- (1520.4045851222957,483.61626524430295) -- (1520.0183759727088,481.53246714482856) -- (1519.7417536754922,479.43131219307463) -- (1519.5754764331348,477.3185595087143) -- (1519.52,475.2) -- (1519.5754764331348,473.08144049128566) -- (1519.7417536754922,470.96868780692535) -- (1520.0183759727088,468.8675328551715) -- (1520.4045851222957,466.7837347556971) -- (1520.8993225518186,464.72300505424994) -- (1521.5012322203722,462.6909920677022) -- (1522.2086643353932,460.6932654024063) -- (1523.0196798746276,458.7353006882916) -- (1523.932055900855,456.8224645705432) -- (1524.943291654806,454.96) -- (1526.0506154095692,453.1530118625917) -- (1527.250992067702,451.40645298720074) -- (1528.5411314802218,449.7251105703026) -- (1529.917497464675,448.1135930545934) -- (1531.3763174975686,446.5763174975686) -- (1532.9135930545933,445.1174974646751) -- (1534.5251105703026,443.74113148022184) -- (1536.2064529872007,442.45099206770215) -- (1537.9530118625917,441.2506154095692) -- (1539.76,440.1432916548059) -- (1541.6224645705431,439.1320559008549) -- (1543.5353006882915,438.21967987462756) -- (1545.4932654024062,437.4086643353933) -- (1547.490992067702,436.70123222037216) -- (1549.52300505425,436.0993225518185) -- (1551.5837347556972,435.6045851222956) -- (1553.6675328551714,435.2183759727088) -- (1555.7686878069253,434.9417536754923) -- (1557.8814404912857,434.7754764331348) -- (1560.0,434.72) -- (1562.1185595087143,434.7754764331348) -- (1564.2313121930747,434.9417536754923) -- (1566.3324671448286,435.2183759727088) -- (1568.4162652443028,435.6045851222956) -- (1570.47699494575,436.0993225518185) -- (1572.509007932298,436.70123222037216) -- (1574.5067345975938,437.40866433539327) -- (1576.4646993117083,438.2196798746275) -- (1578.3775354294569,439.1320559008549) -- (1580.24,440.1432916548059) -- (1582.0469881374083,441.2506154095692) -- (1583.7935470127993,442.45099206770215) -- (1585.4748894296974,443.7411314802218) -- (1587.0864069454067,445.1174974646751) -- (1588.6236825024314,446.57631749756854) -- (1590.082502535325,448.1135930545934) -- (1591.4588685197782,449.7251105703026) -- (1592.749007932298,451.40645298720074) -- (1593.9493845904308,453.1530118625917) -- (1595.056708345194,454.96) -- (1596.067944099145,456.8224645705431) -- (1596.9803201253726,458.7353006882916) -- (1597.7913356646068,460.6932654024062) -- (1598.4987677796278,462.6909920677021) -- (1599.1006774481814,464.72300505424994) -- (1599.5954148777043,466.7837347556971) -- (1599.9816240272912,468.8675328551715) -- (1600.2582463245078,470.96868780692535) -- (1600.4245235668652,473.0814404912856) -- (1600.48,475.2);
\draw[draw=gamfColorAEC4D8,line width=0.31pt] (1600.48,475.2) -- (1600.4245235668652,475.7931966624401) -- (1600.2582463245078,476.3847674140609) -- (1599.9816240272912,476.973090800552) -- (1599.5954148777043,477.55655426840485) -- (1599.1006774481814,478.13355858481003) -- (1598.4987677796278,478.70252222104335) -- (1597.7913356646068,479.26188568732624) -- (1596.9803201253724,479.8101158072783) -- (1596.067944099145,480.34570992024794) -- (1595.056708345194,480.8672) -- (1593.9493845904308,481.37315667847434) -- (1592.749007932298,481.8621931635838) -- (1591.4588685197782,482.3329690403153) -- (1590.082502535325,482.7841939447139) -- (1588.6236825024314,483.2146311006808) -- (1587.0864069454067,483.62310070989093) -- (1585.4748894296974,484.0084831855379) -- (1583.7935470127993,484.36972222104345) -- (1582.0469881374083,484.70582768532057) -- (1580.24,485.01587833665434) -- (1578.3775354294569,485.2990243477607) -- (1576.4646993117085,485.5544896351043) -- (1574.5067345975938,485.7815739860899) -- (1572.509007932298,485.9796549782958) -- (1570.47699494575,486.1481896854908) -- (1568.4162652443028,486.28671616575724) -- (1566.3324671448286,486.39485472764153) -- (1564.2313121930747,486.4723089708621) -- (1562.1185595087143,486.51886659872224) -- (1560.0,486.5344) -- (1557.8814404912857,486.51886659872224) -- (1555.7686878069253,486.4723089708621) -- (1553.6675328551714,486.39485472764153) -- (1551.5837347556972,486.28671616575724) -- (1549.52300505425,486.1481896854908) -- (1547.490992067702,485.9796549782958) -- (1545.4932654024062,485.7815739860899) -- (1543.5353006882915,485.5544896351043) -- (1541.6224645705431,485.2990243477607) -- (1539.76,485.01587833665434) -- (1537.9530118625917,484.70582768532057) -- (1536.2064529872007,484.36972222104345) -- (1534.5251105703026,484.0084831855379) -- (1532.9135930545935,483.62310070989093) -- (1531.3763174975686,483.2146311006808) -- (1529.917497464675,482.7841939447139) -- (1528.5411314802218,482.3329690403153) -- (1527.250992067702,481.8621931635838) -- (1526.0506154095692,481.37315667847434) -- (1524.943291654806,480.8672) -- (1523.932055900855,480.34570992024794) -- (1523.0196798746274,479.8101158072783) -- (1522.2086643353932,479.26188568732624) -- (1521.5012322203722,478.70252222104335) -- (1520.8993225518186,478.13355858481003) -- (1520.4045851222957,477.55655426840485) -- (1520.0183759727088,476.973090800552) -- (1519.7417536754922,476.3847674140609) -- (1519.5754764331348,475.7931966624401) -- (1519.52,475.2) -- (1519.5754764331348,474.60680333755994) -- (1519.7417536754922,474.0152325859391) -- (1520.0183759727088,473.426909199448) -- (1520.4045851222957,472.8434457315952) -- (1520.8993225518186,472.26644141518995) -- (1521.5012322203722,471.6974777789566) -- (1522.2086643353932,471.13811431267374) -- (1523.0196798746276,470.5898841927217) -- (1523.932055900855,470.0542900797521) -- (1524.943291654806,469.53280000000007) -- (1526.0506154095692,469.0268433215257) -- (1527.250992067702,468.5378068364162) -- (1528.5411314802218,468.06703095968476) -- (1529.917497464675,467.61580605528616) -- (1531.3763174975686,467.1853688993192) -- (1532.9135930545933,466.77689929010904) -- (1534.5251105703026,466.39151681446214) -- (1536.2064529872007,466.0302777789566) -- (1537.9530118625917,465.6941723146794) -- (1539.76,465.3841216633457) -- (1541.6224645705431,465.10097565223936) -- (1543.5353006882915,464.8455103648957) -- (1545.4932654024062,464.61842601391015) -- (1547.490992067702,464.42034502170424) -- (1549.52300505425,464.2518103145092) -- (1551.5837347556972,464.1132838342428) -- (1553.6675328551714,464.0051452723585) -- (1555.7686878069253,463.92769102913786) -- (1557.8814404912857,463.88113340127774) -- (1560.0,463.86560000000003) -- (1562.1185595087143,463.88113340127774) -- (1564.2313121930747,463.92769102913786) -- (1566.3324671448286,464.0051452723585) -- (1568.4162652443028,464.1132838342428) -- (1570.47699494575,464.2518103145092) -- (1572.509007932298,464.42034502170424) -- (1574.5067345975938,464.61842601391015) -- (1576.4646993117083,464.8455103648957) -- (1578.3775354294569,465.10097565223936) -- (1580.24,465.3841216633457) -- (1582.0469881374083,465.6941723146794) -- (1583.7935470127993,466.0302777789566) -- (1585.4748894296974,466.39151681446214) -- (1587.0864069454067,466.77689929010904) -- (1588.6236825024314,467.1853688993192) -- (1590.082502535325,467.61580605528616) -- (1591.4588685197782,468.06703095968476) -- (1592.749007932298,468.5378068364162) -- (1593.9493845904308,469.0268433215257) -- (1595.056708345194,469.53280000000007) -- (1596.067944099145,470.0542900797521) -- (1596.9803201253726,470.5898841927217) -- (1597.7913356646068,471.13811431267374) -- (1598.4987677796278,471.6974777789566) -- (1599.1006774481814,472.26644141518995) -- (1599.5954148777043,472.8434457315952) -- (1599.9816240272912,473.426909199448) -- (1600.2582463245078,474.0152325859391) -- (1600.4245235668652,474.60680333755994) -- (1600.48,475.2);
\draw[draw=gamfColorAEC4D8,line width=0.31pt] (1572.9536,475.2) -- (1572.935847541397,477.31855950871443) -- (1572.8826388238424,479.43131219307463) -- (1572.7941196887332,481.53246714482856) -- (1572.6705327608654,483.61626524430295) -- (1572.5122167834181,485.6769949457501) -- (1572.3196056894808,487.70900793229794) -- (1572.0932274126742,489.7067345975938) -- (1571.8337024401192,491.6646993117084) -- (1571.5417421117265,493.5775354294568) -- (1571.2181466704621,495.44) -- (1570.8638030689378,497.2469881374083) -- (1570.4796825383353,498.9935470127993) -- (1570.066837926329,500.6748894296975) -- (1569.626400811304,502.28640694540655) -- (1569.1595784007782,503.8236825024315) -- (1568.66765022253,505.28250253532497) -- (1568.1519646175032,506.65886851977814) -- (1567.6139350440958,507.94900793229795) -- (1567.0550362039708,509.1493845904307) -- (1566.4768,510.2567083451941) -- (1565.8808113374262,511.26794409914515) -- (1565.2687037797466,512.1803201253724) -- (1564.64215507123,512.9913356646067) -- (1564.0028825383354,513.6987677796278) -- (1563.35263838264,514.3006774481814) -- (1562.6932048781769,514.7954148777045) -- (1562.0263894863451,515.1816240272912) -- (1561.3540199017839,515.4582463245076) -- (1560.6779390427887,515.6245235668652) -- (1560.0,515.68) -- (1559.3220609572113,515.6245235668652) -- (1558.6459800982161,515.4582463245076) -- (1557.9736105136549,515.1816240272912) -- (1557.3067951218231,514.7954148777045) -- (1556.64736161736,514.3006774481814) -- (1555.9971174616646,513.6987677796279) -- (1555.35784492877,512.9913356646067) -- (1554.7312962202534,512.1803201253724) -- (1554.1191886625738,511.26794409914515) -- (1553.5232,510.2567083451941) -- (1552.9449637960292,509.1493845904307) -- (1552.3860649559042,507.94900793229795) -- (1551.8480353824968,506.65886851977814) -- (1551.33234977747,505.28250253532497) -- (1550.8404215992218,503.8236825024315) -- (1550.373599188696,502.28640694540655) -- (1549.933162073671,500.6748894296975) -- (1549.5203174616647,498.9935470127993) -- (1549.1361969310622,497.2469881374083) -- (1548.7818533295379,495.44) -- (1548.4582578882735,493.57753542945693) -- (1548.1662975598808,491.6646993117084) -- (1547.9067725873258,489.7067345975938) -- (1547.6803943105192,487.70900793229794) -- (1547.4877832165819,485.6769949457501) -- (1547.3294672391346,483.61626524430295) -- (1547.2058803112668,481.53246714482856) -- (1547.1173611761576,479.43131219307463) -- (1547.064152458603,477.3185595087143) -- (1547.0464,475.2) -- (1547.064152458603,473.08144049128566) -- (1547.1173611761576,470.96868780692535) -- (1547.2058803112668,468.8675328551715) -- (1547.3294672391346,466.7837347556971) -- (1547.4877832165819,464.72300505424994) -- (1547.6803943105192,462.6909920677022) -- (1547.9067725873258,460.6932654024063) -- (1548.1662975598808,458.7353006882916) -- (1548.4582578882735,456.8224645705432) -- (1548.7818533295379,454.96) -- (1549.1361969310622,453.1530118625917) -- (1549.5203174616647,451.40645298720074) -- (1549.933162073671,449.7251105703026) -- (1550.373599188696,448.1135930545934) -- (1550.8404215992218,446.5763174975686) -- (1551.33234977747,445.1174974646751) -- (1551.8480353824968,443.74113148022184) -- (1552.3860649559042,442.45099206770215) -- (1552.9449637960292,441.2506154095692) -- (1553.5232,440.1432916548059) -- (1554.1191886625738,439.1320559008549) -- (1554.7312962202534,438.21967987462756) -- (1555.35784492877,437.4086643353933) -- (1555.9971174616646,436.70123222037216) -- (1556.64736161736,436.0993225518185) -- (1557.3067951218231,435.6045851222956) -- (1557.9736105136549,435.2183759727088) -- (1558.6459800982161,434.9417536754923) -- (1559.3220609572113,434.7754764331348) -- (1560.0,434.72) -- (1560.6779390427887,434.7754764331348) -- (1561.3540199017839,434.9417536754923) -- (1562.0263894863451,435.2183759727088) -- (1562.6932048781769,435.6045851222956) -- (1563.35263838264,436.0993225518185) -- (1564.0028825383354,436.70123222037216) -- (1564.64215507123,437.40866433539327) -- (1565.2687037797466,438.2196798746275) -- (1565.8808113374262,439.1320559008549) -- (1566.4768,440.1432916548059) -- (1567.0550362039708,441.2506154095692) -- (1567.6139350440958,442.45099206770215) -- (1568.1519646175032,443.7411314802218) -- (1568.66765022253,445.1174974646751) -- (1569.1595784007782,446.57631749756854) -- (1569.626400811304,448.1135930545934) -- (1570.066837926329,449.7251105703026) -- (1570.4796825383353,451.40645298720074) -- (1570.8638030689378,453.1530118625917) -- (1571.2181466704621,454.96) -- (1571.5417421117265,456.8224645705431) -- (1571.8337024401192,458.7353006882916) -- (1572.0932274126742,460.6932654024062) -- (1572.3196056894808,462.6909920677021) -- (1572.5122167834181,464.72300505424994) -- (1572.6705327608654,466.7837347556971) -- (1572.7941196887332,468.8675328551715) -- (1572.8826388238424,470.96868780692535) -- (1572.935847541397,473.0814404912856) -- (1572.9536,475.2);
\draw[->,draw=gamfColor7056AC,line width=0.80pt] (1560,475.2) -- (1592.56,440.0);
\fill[fill=gamfColor18334B] (1560,475.2) circle[radius=0.038cm];
\node[anchor=center,align=center,text=gamfColor7056AC,font=\sffamily\fontsize{7.10}{8.66}\selectfont,inner sep=0pt] at (1601.36,433.84) {$\hat{\mathbf u}$};
\end{tikzpicture}
}
\caption{GAMF framework. Shared audio and ATF encoders form observation queries and candidate keys, which are matched at corresponding microphones and aggregated into direction scores. GAMF-L decodes learned scores; GAMF-H combines a gated learned contribution with analytical scores adapted from GSRP-NMF-frob~\cite{muller2025gsrp}. $M$ denotes active microphones; $T=33$, $F=129$, and $K=384$. Layer details are given in the text.}
\label{fig:main}
\end{figure*}
\section{Introduction}
Direction-of-arrival (DOA) estimation supports robotics, teleconferencing, and spatial audio. Classical approaches include generalized cross-correlation~\cite{knapp1976}, MUSIC~\cite{schmidt1986}, and SRP-PHAT~\cite{dibiase2001,grinstein2024review}. Neural methods learn spatial features from multichannel signals~\cite{grumiaux2022,adavanne2018,chakrabarty2019}. SRP-DNN estimates direct-path phase differences before spatial matching~\cite{yang2022srpdnn}, while Neural-SRP embeds SRP-inspired processing in a learned architecture~\cite{grinstein2024neuralsrp}. These approaches demonstrate the value of combining learning with physical spatial cues. Deploying one model across devices, however, requires robustness to changes in microphone layout and count~\cite{schwartz2023}.

Microphone-pair training combines time-delay learning with geometry-aware DOA training~\cite{an2024pair}; IPDnet supports variable arrays through direct-path phase-difference estimation~\cite{wang2024ipdnet}. Other networks explicitly encode microphone coordinates~\cite{kowalk2023}. GI-DOAEnet introduces microphone positional encoding for geometry-invariant estimation~\cite{baek2025}. PhaseCoder adopts its phase-modulation encoding in a Transformer spatial encoder and extends the representations to audio-language tasks~\cite{phasecoder}. Coordinates nevertheless only partially describe array acoustics: device shadowing, acoustic ports, and microphone directivity alter interchannel amplitude and phase relationships beyond free-field predictions. Although training may capture such effects, coordinates do not explicitly specify them for a new array. Directional array transfer functions (ATFs), obtained through modeling or measurement, provide richer descriptors by associating candidate directions with complex microphone responses.

ATFs support both analytical and learned localization. They serve as steering templates in MUSIC~\cite{suzuki2017,hark}, while generalized steered response power (GSRP) incorporates generic acoustic models and jointly exploits interchannel level and phase differences~\cite{muller2025gsrp}. Its normalized matched-filter formulation retains these cues while normalizing spectral power. Very recently, Heikkinen et al.\ introduced separate signal and ATF encoders with cross-attention, demonstrating generalization to unseen arrays with the same microphone count~\cite{heikkinen2026}. Extending ATF conditioning to varying microphone counts remains important for flexible deployment.

Building on GI-DOAEnet and PhaseCoder~\cite{baek2025,phasecoder}, we propose a Generalizable ATF Matching Framework (GAMF) for three-dimensional DOA estimation across array geometries and microphone counts. Its central extension enriches geometry-conditioned learning with directional ATF embeddings. Local spectral features and multichannel temporal context form observation queries; ATF and direction embeddings form candidate keys. Matching at corresponding microphones preserves each channel's association with its acoustic response before aggregation. Shared encoders and masked aggregation accommodate variable microphone counts.

The framework provides standalone learned matching (GAMF-L) and hybrid matching (GAMF-H). The latter gates the learned contribution and combines it with analytical scores adapted from GSRP-NMF-frob~\cite{muller2025gsrp}. Experiments span 3--8 microphones, source distances, noise, and reverberation, together with LOCATA Task~1 recordings~\cite{evers2020}. Both configurations outperform the retrained positional-encoding baseline in all six synthetic conditions. The hybrid combines learned robustness under noise and reverberation with the strength of analytical signal processing in favorable scenes, yielding competitive performance across both regimes. On eight-microphone LOCATA, its mean errors are $3.76\degree$ for three-dimensional DOA and $2.87\degree$ for azimuth. The code and pretrained model will be released publicly soon.

\section{Method}
The proposed framework estimates DOA by matching multichannel observations to candidate directional ATFs (Fig.~\ref{fig:main}). We propose two configurations: ATF-conditioned learned matching (GAMF-L) and hybrid matching (GAMF-H), which integrates learned scores with an analytical branch adapted from GSRP-NMF-frob~\cite{muller2025gsrp}. The analytical branch is also evaluated separately as an ablation reference. The learned formulation performs contextual matching in a feature space, while the analytical formulation evaluates spectral agreement using the same candidate library.

Using the $e^{-\mathrm{i}2\pi ft}$ Fourier convention for both STFTs and ATFs, let $X_{j,t,f}$ denote the complex STFT coefficient at microphone $j$, frame $t$, and frequency bin $f$, with $M$ active microphones, $T$ frames, and $F$ frequency bins. Let $\mathbf p=\{\mathbf p_j\}_{j=1}^{M}$ denote the microphone coordinates relative to their centroid, with $\mathbf p_j\in\mathbb R^3$. Each candidate unit direction $\mathbf u_k$ ($k=1,\ldots,K$) references this centroid and has a directional ATF $H_{j,k,f}$. Before encoding, observation and ATF vectors are normalized to unit norm across active microphones, yielding $\bar{\mathbf x}_{t,f}$ and $\bar{\mathbf h}_{k,f}$ while preserving relative interchannel amplitudes and phases. Near-zero vectors are invalidated; padding is masked in attention and aggregation.
\subsection{ATF-conditioned learned matching}
Figure~\ref{fig:main} summarizes the learned path: observation-query encoding, candidate-ATF key encoding, and microphone-aligned scoring. Both microphone and candidate-direction axes are retained until matching.

\subsubsection{Observation queries}
For each microphone and frame, the full normalized spectrum is represented by $\mathbf a_{j,t}=[\operatorname{Re}\bar X_{j,t,:};\operatorname{Im}\bar X_{j,t,:}]\in\mathbb R^{2F}$. A shared spectral encoder $E_x$, consisting of linear $2F\!\to\!D$, layer normalization, GELU, and linear $D\!\to\!D$ layers, produces $\mathbf z_{j,t}=E_x(\mathbf a_{j,t})$. Its weights are shared across microphones and frames, producing an observation tensor of size $M\times T\times D$.

A $D\!\to\!D$ affine projection of $\mathbf z_{j,t}$ supplies local spectral information. A parallel context path adds positional encoding derived from $\mathbf p_j$, time encoding, and sequence encoding before flattening the microphone--frame grid in frame-major order. The microphone encoding follows GI-DOAEnet and its adoption in PhaseCoder~\cite{baek2025,phasecoder}. Masked pre-normalization Transformer blocks~\cite{vaswani2017} model dependencies across microphones and time, with multi-head self-attention, a GELU feed-forward network, and residual connections. Three blocks with four attention heads and feed-forward widths 384--512--384 restore contextual features $\mathbf c_{j,t}$ on the microphone--frame grid. For head $\ell$, the query is
\begin{equation}
\mathbf q_{j,t}^{(\ell)}=\operatorname{norm}\!\left(
W_{\rm loc}^{(\ell)}\mathbf z_{j,t}
+\beta W_{\rm ctx}^{(\ell)}\mathbf c_{j,t}\right),
\label{eq:query}
\end{equation}
Here $D$ is the embedding dimension, $\beta>0$ is learned, projections include biases, and $\operatorname{norm}$ denotes unit-norm scaling. Local and contextual projections are combined, split into four 96-dimensional heads, and normalized per head, yielding $M\times T\times4\times96$ queries.

\subsubsection{Candidate ATF keys}
For each microphone and candidate, we similarly form $\mathbf b_{j,k}=[\operatorname{Re}\bar H_{j,k,:};\operatorname{Im}\bar H_{j,k,:}]$. A shared ATF encoder $E_h$ has the same layer structure as $E_x$ but independent weights. A direction encoder with linear $3\!\to\!D$, ReLU, and linear $D\!\to\!D$ layers embeds the candidate unit direction as $E_u(\mathbf u_k)$. The direction embedding is broadcast over microphones and added to ATF features. Layer normalization, microphone masking, and a $D\!\to\!D$ projection precede head splitting and normalization:
\begin{equation}
\mathbf k_{j,k}^{(\ell)}=\operatorname{norm}\!\left[
W_{\rm key}^{(\ell)}\operatorname{LN}\!\left(
E_h(\mathbf b_{j,k})+E_u(\mathbf u_k)\right)\right].
\label{eq:atf_key}
\end{equation}
Here $E_h$ encodes the array response and $E_u$ its direction. The resulting $M\times K\times4\times96$ keys are reused across frames. Encoder weights are shared across microphones and candidates, with no candidate-specific output weights. Coordinates condition observation context; ATFs encode acoustics in the keys.

\subsubsection{Microphone-aligned direction scoring}
Each head compares its query and key at the same microphone, producing an $M\times T\times K\times4$ tensor before microphone aggregation. For active-channel mask $m_j$ and $M=\sum_j m_j$, the head score and learned direction score are
\begin{equation}
\begin{aligned}
r_{t,k}^{(\ell)}&=\frac{1}{M\sqrt d}\sum_j m_j
(\mathbf q_{j,t}^{(\ell)})^{\mathsf T}\mathbf k_{j,k}^{(\ell)},\\
S^{\rm learn}_{t,k}&=\tau_l\left(\mathbf w^{\mathsf T}\mathbf r_{t,k}+b\right),
\end{aligned}
\label{eq:learned_matching}
\end{equation}
where $d=D/L$ with $L=4$ heads, $\mathbf r_{t,k}$ collects their scores, and $\tau_l>0$ is a learned scale. Masked microphone averaging yields $T\times K\times4$ scores, followed by a learned linear $4\!\to\!1$ merger and positive scaling to produce the $T\times K$ direction map, without value-vector aggregation. Shared encoders, padded-token masking, and active-channel score averaging allow one parameter set to support variable microphone counts within the configured limit.

For valid frames $\mathcal T$, GAMF-L uses the learned score map alone:
\begin{equation}
s_k^{\rm L}=\frac{1}{|\mathcal T|}\sum_{t\in\mathcal T}S^{\rm learn}_{t,k},
\qquad \hat{\mathbf u}_{\rm L}=\mathbf u_{\arg\max_k s_k^{\rm L}}.
\label{eq:learned_decode}
\end{equation}
GAMF-H combines learned and analytical scores over the same candidates; both configurations connect array acoustics to DOA through candidate-dependent ATF keys.

\subsection{Analytical ATF matching}
The analytical branch adapts GSRP-NMF-frob~\cite{muller2025gsrp} using instantaneous covariance and valid-frequency averaging, preserving the normalized matching criterion while aligning with framewise learned scores. For valid frequency bins $\mathcal F_{t,k}$, its score is
\begin{equation}
S^{\rm phys}_{t,k}=\frac{\sum_{f\in\mathcal F_{t,k}}
|\bar{\mathbf h}_{k,f}^{\mathsf H}\bar{\mathbf x}_{t,f}|^2}
{\max(1,|\mathcal F_{t,k}|)}.
\label{eq:physical}
\end{equation}
With instantaneous covariance $\mathbf R_{t,f}=\mathbf x_{t,f}\mathbf x_{t,f}^{\mathsf H}$, the per-frequency term equals NMF-frob~\cite[Eq.~(42)]{muller2025gsrp}, directly measuring spectral agreement between observations and candidate ATFs. This criterion serves as GAMF-H's analytical component and an ablation reference. The branches share ATFs and candidate directions, using analytical matching and learned representations, respectively.

\subsection{Integration and training}
In GAMF-H, an MLP with sigmoid output computes a framewise gate $g_t\in(0,1)$. Its inputs describe the concentration and temporal stability of analytical scores, disagreement between branch distributions, signal energy, valid-frequency fraction, and active-microphone count. It modulates the learned contribution before averaging over valid frames $\mathcal T$:
\begin{equation}
s_k=\frac{1}{|\mathcal T|}\sum_{t\in\mathcal T}
\left[\tau_p S^{\rm phys}_{t,k}+\alpha g_t S^{\rm learn}_{t,k}\right].
\label{eq:fusion}
\end{equation}
Here $\tau_p,\alpha>0$ are learned scales. The gate scales the learned contribution while retaining the analytical score; it does not form a convex average of the branches. The output is $\hat{\mathbf u}=\mathbf u_{\arg\max_k s_k}$.

When training GAMF-H, soft labels $q_k\propto\exp(80\mathbf u_k^{\mathsf T}\mathbf u)$ for target-speech direction $\mathbf u$ supervise the temporally averaged fused, physical, and learned scores through cross-entropy~\cite{perotin2019}, with weights $1$, $0.2$, and $0.1$. An auxiliary speech-presence head takes masked averages of contextual features and mean log energy and uses unit-weight binary cross-entropy on all samples. GAMF-L is trained independently using learned direction scores and the auxiliary speech-presence task. All learned positive scales use softplus.

\begin{table*}[t]
\centering
\caption{Synthetic localization averaged over evaluations with 3, 4, 5, 6, 7, and 8 microphones. The left and right column groups represent relatively easy (low reverberation) and challenging (medium reverberation with more severe noisy conditions) scenarios, respectively. Entries report spherical MAE in degrees / Acc@10 in percent; the two best values for each metric within a column are bold.}
\label{tab:synthetic}
\setlength{\tabcolsep}{1.4pt}
\begin{tabular*}{\textwidth}{@{\extracolsep{\fill}}lrrrrrr@{}}\toprule
 & \multicolumn{3}{c}{Low reverberation (0.08--0.25 s)} & \multicolumn{3}{c}{Medium reverberation (0.25--0.80 s)}\\
\cmidrule(lr){2-4}\cmidrule(l){5-7}
Method & Clean & 15 dB & 10 dB & Clean & 0 dB & $-10$ dB\\
 & MAE/Acc@10 & MAE/Acc@10 & MAE/Acc@10 & MAE/Acc@10 & MAE/Acc@10 & MAE/Acc@10\\
\midrule
MUSIC-ATF~\cite{suzuki2017,hark} & 14.11/66.10 & 15.85/63.60 & 18.65/59.07 & 32.39/35.78 & 45.10/24.07 & 62.81/12.55\\
SRP-PHAT~\cite{dibiase2001} & 13.38/62.13 & 21.45/49.97 & 28.40/42.40 & 22.45/48.62 & 46.11/25.03 & 60.02/13.62\\
GSRP-NMF-frob~\cite{muller2025gsrp} & \textbf{7.73}/\textbf{90.17} & 16.71/71.53 & 24.04/60.23 & 21.15/64.52 & 45.27/31.90 & 59.79/17.43\\
PhaseCoder (retrained)~\cite{phasecoder} & 13.32/64.37 & 14.84/59.53 & 15.90/56.23 & 18.38/44.18 & 27.23/27.87 & 41.26/15.57\\
\midrule
Analytical branch & \textbf{8.22}/\textbf{90.73} & 19.43/72.43 & 26.03/63.63 & 17.60/\textbf{75.92} & 41.46/42.82 & 55.21/26.55\\
GAMF-L & 8.42/86.03 & \textbf{10.92}/\textbf{79.50} & \textbf{11.75}/\textbf{76.40} & \textbf{11.86}/70.75 & \textbf{20.84}/\textbf{48.27} & \textbf{33.41}/\textbf{31.73}\\
GAMF-H & 8.68/86.53 & \textbf{11.77}/\textbf{79.50} & \textbf{13.13}/\textbf{75.10} & \textbf{12.19}/\textbf{73.10} & \textbf{23.93}/\textbf{48.32} & \textbf{38.00}/\textbf{30.30}\\
\bottomrule
\end{tabular*}\end{table*}

\begin{table*}[t]
\centering
\caption{Six- and eight-microphone LOCATA Task~1 results. Sph., Az., and El. are MAE in degrees. Acc@10 uses spherical error except for GI-DOAEnet-FM (azimuth only, $^*$). Bold marks the two best 3D results; / denotes unavailable metrics.}
\label{tab:locata}
\small
\setlength{\tabcolsep}{2.2pt}
\begin{tabular*}{\textwidth}{@{\extracolsep{\fill}}lrrrrrrrr@{}}

\toprule
 & \multicolumn{4}{c}{6 microphones} & \multicolumn{4}{c}{8 microphones}\\
\cmidrule(lr){2-5}\cmidrule(l){6-9}
Method & \shortstack[r]{Sph.\\MAE$\downarrow$} & \shortstack[r]{Az.\\MAE$\downarrow$} & \shortstack[r]{El.\\MAE$\downarrow$} & Acc@10$\uparrow$ & \shortstack[r]{Sph.\\MAE$\downarrow$} & \shortstack[r]{Az.\\MAE$\downarrow$} & \shortstack[r]{El.\\MAE$\downarrow$} & Acc@10$\uparrow$\\
\midrule
PhaseCoder (retrained)~\cite{phasecoder} & 10.30 & 6.53 & 6.46 & 61.68 & 8.97 & 4.94 & 6.69 & 73.42\\
MUSIC-ATF~\cite{suzuki2017,hark} & 13.85 & 10.24 & 7.47 & 50.24 & 7.48 & 4.18 & 5.51 & 75.95\\
SRP-PHAT~\cite{dibiase2001} & 9.76 & 8.49 & 3.23 & 54.67 & 6.49 & 4.64 & 3.90 & 78.81\\
GSRP-NMF-frob~\cite{muller2025gsrp} & 6.62 & 5.79 & 2.72 & 79.51 & 4.12 & 3.26 & 1.93 & 92.81\\
GI-DOAEnet-FM~\cite{baek2025} & / & 8.86 & / & 87.17$^{*}$ & / & 4.70 & / & 95.12$^{*}$\\
\midrule
Analytical branch & \textbf{5.03} & \textbf{4.12} & \textbf{2.24} & \textbf{95.98} & \textbf{3.87} & \textbf{3.04} & \textbf{1.87} & \textbf{94.14}\\
GAMF-L & 6.97 & 5.40 & 3.58 & 78.79 & 6.09 & 4.55 & 3.31 & 86.68\\
GAMF-H & \textbf{5.23} & \textbf{4.16} & \textbf{2.54} & \textbf{94.10} & \textbf{3.76} & \textbf{2.87} & \textbf{1.86} & \textbf{98.80}\\
\bottomrule
\end{tabular*}
\end{table*}

\section{Experiments and Discussions}
\subsection{Implementation and training}
We process 250-ms, 16-kHz segments padded to 4096 samples. A centered 256-point Hann-window STFT with a 128-sample hop yields 33 frames and 129 frequency bins.

Candidate ATFs use 384 Fibonacci-sphere directions~\cite{swinbank2006} at 1~m from the active-microphone centroid. Free-field propagation models point microphones, while the Duda--Martens model~\cite{duda1998} efficiently and accurately captures rigid-sphere shadowing and scattering. Simulated observations include image-source room reflections and reverberation~\cite{allen1979,jarrett2012}; candidate ATFs omit reflections and retain the fixed reference distance.

Training equally samples 300,000 point-microphone and 300,000 rigid-sphere scenes with 3--8 microphones. Rooms span 3--10.5~m in length and width and 2.5--5~m in height, with target \RT{} of 0.08--1.42~s. Sphere diameters are 7--18~cm with at least 1-cm microphone spacing. LibriSpeech targets~\cite{panayotov2015} and FSD50K interference with human-vocalization-tagged clips excluded~\cite{fonseca2022} are convolved with the simulated responses. GAMF-L and GAMF-H are trained independently using AdamW~\cite{loshchilov2019} and a batch size of 512. The three-stage curriculum uses 120k, 90k, and 20k updates: clean training followed by mixtures of clean and noisy inputs, with noisy-input SNRs of $-5$ to $15$~dB and $-10$ to $15$~dB, respectively. Evaluation uses the third-stage checkpoints at 20k updates. SNR uses segment power across active channels; \emph{Clean} means no added interference, with reverberation retained.

\subsection{Baselines and evaluation}\label{sec:protocol}
Evaluation combines controlled simulations varying SNR, microphone count, source distance, and reverberation with LOCATA recordings to assess real-world transfer.

PhaseCoder Medium~\cite{phasecoder} is retrained on our data; its azimuth/elevation heads are decoded excluding no-source classes. MUSIC-ATF follows HARK's single-source procedure over 500--2687.5~Hz~\cite{schmidt1986,suzuki2017,hark}. SRP-PHAT uses free-field far-field steering over 300--3500~Hz~\cite{dibiase2001}. GSRP-NMF-frob~\cite{muller2025gsrp} uses this band and covariance smoothing, $R_t=0.2R_{t-1}+0.8R_t^{\rm inst}$; our analytical ablation uses GAMF-H's analytical scores with instantaneous covariance and all valid bins. Classical baselines share the direction grid, and ATF baselines share the candidate library.

The 9,000 synthetic scenes use 56--65-mm-radius rigid spheres, covering five distance intervals (0.2--6~m), three \RT{} intervals (0.08--0.8~s), and 3--8 microphones, with 100 scenes per combination. Table~\ref{tab:synthetic} groups these into low (0.08--0.25~s) and medium (0.25--0.80~s) reverberation, containing 3,000 and 6,000 scenes per SNR, respectively. Audio clips are fixed across SNRs.

LOCATA uses 16 Task~1 recordings~\cite{evers2020} with NAO channels 1,2,5,6,7,10; the eight-microphone subset adds channels 3 and 12. Candidate ATFs are synthesized using the Duda--Martens model~\cite{duda1998} with a 57-mm sphere centered at the array origin; DOA labels reference the active-microphone centroid. Three-dimensional methods produce one estimate per nonoverlapping 250-ms segment. The azimuth-only GI-DOAEnet-FM~\cite{baek2025} processes each recording as a whole using a 32-ms STFT window and an 8-ms hop. Speech-activity thresholds are 50\% and 66.66\%, respectively.

Spherical MAE measures angular separation; Acc@10 reports the percentage within $10\degree$. Azimuth errors are wrapped and elevation errors absolute. Synthetic metrics are sample-averaged; LOCATA metrics are averaged within each recording and equally across recordings. GI's Acc@10 uses azimuth rather than spherical error.

\subsection{Synthetic localization}
In Table~\ref{tab:synthetic}, GAMF-L outperforms retrained PhaseCoder in MAE across all six conditions and achieves the lowest MAE in five. GAMF-L and GAMF-H occupy the top two ranks on both metrics in all four noisy conditions. GAMF-L's strong standalone performance shows that learned ATF-conditioned matching is effective independently of analytical fusion. These comparisons do not isolate the contribution of ATF embeddings from other design changes.

GSRP-NMF-frob~\cite{muller2025gsrp} leads in clean, low-reverberation MAE ($7.73\degree$), versus $8.42\degree$ and $8.68\degree$ for GAMF-L and GAMF-H. Conversely, GAMF-H reduces analytical-branch MAE by 31.2\%--49.5\% across noisy conditions. Even without added interference, increasing reverberation favors learned matching: under medium reverberation, GAMF-L and GAMF-H achieve lower MAE than the analytical baselines. Thus, analytical signal processing excels when observations closely follow ATF templates, while learned matching better accommodates interference and reverberation. Their integration is competitive across benign and challenging scenes, although GAMF-L has lower MAE throughout the synthetic tests.

\subsection{Real-recording localization}
On eight-microphone LOCATA (Table~\ref{tab:locata}), GAMF-H leads with $3.76\degree$ spherical MAE, $2.87\degree$ azimuth MAE, and 98.80\% Acc@10. Compared with the analytical branch, its spherical MAE decreases only from $3.87\degree$ to $3.76\degree$, but Acc@10 rises from 94.14\% to 98.80\%, indicating fewer errors above $10\degree$. With six microphones, the analytical branch slightly outperforms GAMF-H. GAMF-L outperforms retrained PhaseCoder but trails analytical matching on both subsets. Static single-source recordings without added interference can favor physical matching when directional cues remain reliable, although reverberation and simulation mismatch also matter. GAMF-H combines analytical accuracy on real recordings with learned robustness in noisy simulations, remaining competitive without always outperforming either branch.

\section{Conclusion}
We presented GAMF for three-dimensional DOA estimation across array geometries and microphone counts. Its central contribution is to enrich geometry-conditioned neural estimation with directional ATFs, explicitly describing array acoustics beyond microphone coordinates. Microphone-aligned matching connects contextual observation features to candidate acoustic responses, while shared encoders and masked aggregation support variable microphone counts. This formulation supports both standalone learned matching (GAMF-L) and its integration with adapted analytical signal processing (GAMF-H). The experiments reveal complementary strengths rather than a uniformly superior matching strategy. Learned matching is effective under noise and reverberation, whereas analytical matching remains strong when directional acoustic cues are reliable. Combining them yields competitive performance across these conditions, including transfer from simulated ATFs to real LOCATA recordings. These findings support ATF-conditioned learning as a useful extension of positional encoding and motivate adaptive integration with physical matching. Future work will isolate ATF-conditioning benefits, test measured ATFs and new geometries, and improve mismatch robustness.

\clearpage
\section*{Acknowledgments}
This work was supported in part by the Science and Technology Program of Jiangsu Province under Grant BZ2024062.

\end{document}